# A Similarity Measure for Weaving Patterns in Textiles

Sven Helmer Vuong M. Ngo

## ABSTRACT

We propose a novel approach for measuring the similarity between weaving patterns that can provide similarity-based search functionality for textile archives. We represent textile structures using hypergraphs and extract multisets of *k*-neighborhoods from these graphs. The resulting multisets are then compared using Jaccard coefficients, Hamming distances, and cosine measures. We evaluate the different variants of our similarity measure experimentally, showing that it can be implemented efficiently and illustrating its quality using it to cluster and query a data set containing more than a thousand textile samples.

## 1. INTRODUCTION

Textiles and their creation play an important part in studying history and prehistory. In fact, some cultures, such as the pre-Columbian civilizations found in the Andes, documented and communicated complex information via textiles [27]. Although today many fabrics are produced using fully automated mechanical looms, traditional methods of weaving are still kept alive in communities around the world, including countries as diverse as the United Arab Emirates, China, and Peru. Compared to industrially produced textiles, archaeological specimens and artifacts created manually by traditional craftsmen exhibit a significantly more complex structure, which makes it much harder to represent, compare, and retrieve them in an information system.

In the humanities, the digitization of cultural heritage and cultural practices plays an ever more important role [31]. When looking at publicly accessible digital archives for textiles, we encounter the following situation. The American Textile History Museum (`http://www.athm.org/`) and the TEXMEDIN digital library (`http://www.texmedindigitallibrary.eu/`) offer keyword search over their collections, while the Textile Museum of Canada (`http://www.textilemuseum.ca/`) also supports browsing facilities according to different categories, such as textile type, region, materials, techniques, and period, and the University of Leeds International Textile Archive (ULITA) (`http://ulita.leeds.ac.uk/`) organizes their collection by region. In an earlier project [24], "Weaving Communities of Practice", we provided additional search functionality by applying an ontological approach to building a knowledge base for Andean textiles. However, this system, like the other archives mentioned above, does not offer a facility to search and compare textiles according to their internal structure.

Providing techniques to represent and compare textiles according to their structure would help researchers investigating them (such as archaeologists, ethnographers, and anthropologists) to search for variations of specific patterns. In turn this facilitates the process of identifying fabrics and the methods used to create them. Moreover, this can help domain experts gain deeper insights into how textiles evolved over time and spread across different regions. With our work, we contribute to this effort in the following ways.

- We propose a novel approach for modeling textiles using hypergraphs that can represent many different structures, such as woven, knitted, or braided textiles. Orientation invariance is an important feature of our representation.
- Based on this representation we develop similarity measures to compare textiles at the structural level.
- We illustrate the effectiveness of our technique in an experimental evaluation clustering a data set containing over a thousand textiles, showing that our clustering comes close to the correct categorization. Additionally, we ran experiments querying our text collection and achieved very good results in terms of retrieval performance, e.g., a mean average precision (MAP) of around 0.93.

The remainder of the paper is organized as follows. In the next section we review the related work and Section 3 covers existing methods for modeling textile structures in more detail and discusses their merits and drawbacks. We introduce our approach in Section 4 and its implementation in Section 5. An experimental evaluation and its results are presented in Section 6. A summary and outlook concludes the paper.

## 2. RELATED WORK

One of the most comprehensive and systematic expositions on classifying the structures of textiles according to basic patterns was compiled by Emery [9], who was a curator at the Textile Museum in Washington D.C. Although Emery's terminology is widely used, the use of terms is not always consistent, even among domain experts [4]. This is especially true for textiles produced in different cultural contexts such as the South American Andes [1, 8, 32], specimens of which we were confronted with in our earlier project [24]. Additionally, due to the huge diversity of textile structures, it is impractical to come up with a complete natural language classification of every possible structure. Consequently, there have been several attempts to define textile structures in a formal and mathematical way [12, 13].

Textiles created by mechanical looms exhibit a very regular pattern and can be described using a grid structure, which can be mapped to a binary matrix representation [17, 25]. Since we set out to also describe structures originating from manual techniques, whose shape may be much more irregular, this is not an option for us. A more general approach for textile modeling is based on topology, i.e., representing fabrics with the help of elements taken from knot theory. Early research in this area specialized on specific techniques, such as knitting [20, 21] and is therefore not generally applicable. Grishanov et al. [12, 13] developed a more general method relying on tangles, which are knot fragments embedding arcs into a sphere [6].[1] While this approach is more expressive compared to previous topological models, it is only applicable to structures that are periodical in two perpendicular directions. Also, certain structures that can only be created manually and multi-layered disjoint fabrics are not considered. Furthermore, the main purpose of the topological techniques described above are the enumeration and classification of textile structures, not their fast retrieval. For instance, the problem of determining whether two given structures described by knots, links, or tangles are equivalent is intractable in the general case [7].

Another way to represent textiles consists of specifying structures based on image-processing techniques. Most of these techniques, e.g. those described in [26, 33], can only be applied to specific textile structures, such as regular rectangular grids or knitting. Moreover, while image processing has the advantage that most of the preprocessing can be automated, textiles, especially complex ones, are three-dimensional objects and some of the structure may be hidden. Ma et al. introduce a special binary encoding to retain some of this information [22], however, their approach is also only applicable to regular grid-like structures. From our point of view Zheng et al. come closest to what we have in mind by developing a method for indexing and retrieving textiles based on their structure [36]. Their technique, though, is only applicable to plain, twill, and satin weave patterns and also shares most of the drawbacks of the other image-processing approaches.

# 3. TEXTILE MODELING

Before going into details about existing techniques for modeling textiles we have to define the term *textile structure*, by which we mean the spatial relationships between elements such as yarns, threads, strands, or other similar long and continuous spun pieces of fiber. Basically, given (part of) a textile structure we want to be able to find other structures in which the elements are arranged in a similar manner. In her book [9], Emery distinguishes three broad types of textile structures: interworked elements, interlaced elements, and felted fibers. The last one, felted fiber, is not of interest to us, as it is created by compressing, matting, and condensing fibers. This results in the fibers getting entangled with each other in a very irregular and dense pattern, which makes it virtually impossible, not to mention irrelevant, to identify all of the relationships between individual fibers. In interworked elements, threads are connected by means of knotting, linking, stitching, looping, or twining, whereas in interlaced elements, they pass over and under each other without connecting in any other way. Providing all the details of Emery's full classification scheme goes beyond the scope of this paper, due to the fact that our goal is to develop a formal mathematical model of textile structures rather than establishing natural language descriptions. Nevertheless, we show some typical examples of textile structures in Figure 1 (Fig. 1(a) and (b) illustrate interworked elements, Fig. 1(c) and (d) interlaced ones).

[1]We provide more background on knots, links, and tangles in Section 3.

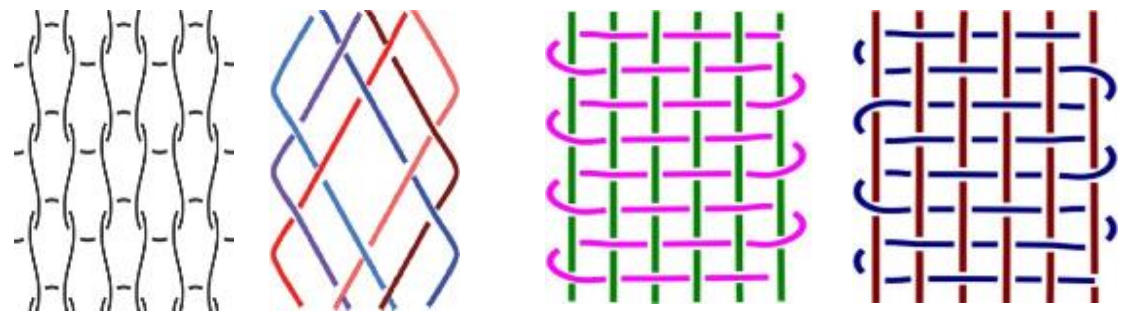

(a) Knitting (b) Braid (c) Plain weave (d) Twill (2/1)

Interworked elements Interlaced elements

**Figure 1: Examples of textile structures**

In the following we give a brief informal summary of topological concepts that have been used to model textile structures, in particular *knots*, *links*, *tangles*, and their two-dimensional projections. Detailed, formal definitions can be found in [7, 12].

## 3.1 Knots

A *knot* is a one-dimensional subset of points $K \subset \mathbb{R}^3$ homeomorphic to a circle. Figure 2(a) shows a trivial knot, a circle, while Figure 2(b) depicts a so-called trefoil. We want to be able to distinguish the different types of knots, i.e., determine whether two knots are equivalent. Intuitively, two knots are of the same type if we can continuously deform on knot into the other without breaking it or intersecting it with itself. Figure 2(c) and (d) shows an example of two knots that can be transformed into each other.

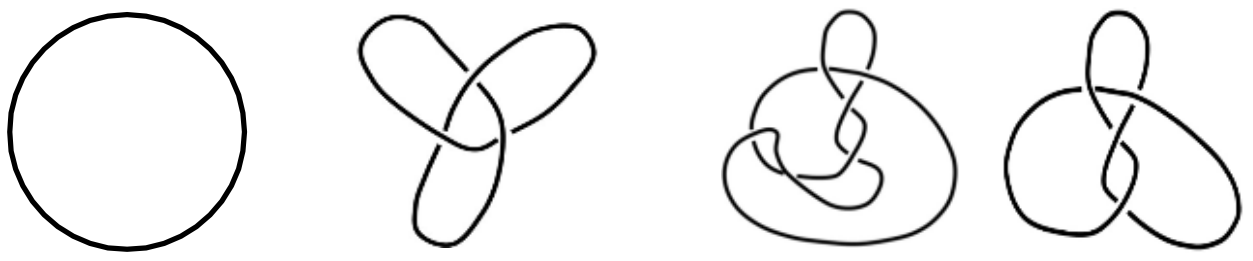

(a) Trivial knot (b) Non-trivial knot (c) Knot A (d) Knot B

Types Transformations

**Figure 2: Examples of knots and transformations**

## 3.2 Links

A knot is an embedding of a single circle into $\mathbb{R}^3$, this can be generalized to a collection of knots, which is called a *link*. Figure 3 shows two link examples, a trivial link in (a) and Borromean rings in (b). Similar to knots, two links are equivalent if we can transform one into the other by deforming it without cutting it.

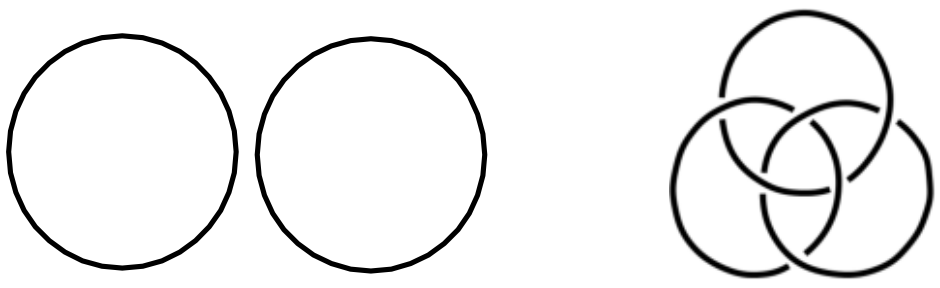

(a) Trivial link (b) Borromean rings

**Figure 3: Examples of links**

## 3.3 Tangles

While the theory of knots and links provides a well-established foundation and mathematical tools, it cannot be directly applied to textile structures. Rarely are textiles made up of a collection of intertwined and deformed circles. Conway introduced knot fragments called *tangles* [6], which can be used to describe knots and links as well as textile structures. Conway's original intention was to develop a simpler notation for systematically enumerating and classifying knots and links. Grishanov et al. first applied this concept to describing textile structures [11].

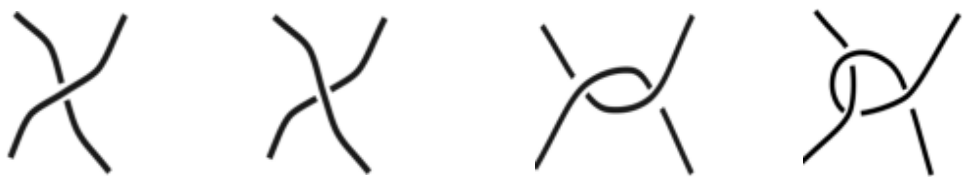

**Figure 4: Examples of tangles**

In his original definition, Conway described a tangle as a knot fragment with two arcs protruding into the four corners of the tangle. The corners are labeled NW, NE, SW, and SE after a compass rose (see Figure 4 for examples). Conway's definition of tangles has also been generalized to *n-tangles* containing *n* arcs instead of just two.

## 3.4 Two-dimensional Projections

As the main interest is usually in the type of a structure and not its exact three-dimensional form, often a planar representation is applied, projecting knots and links from $\mathbb{R}^3$ to $\mathbb{R}^2$. While this loses knowledge of the exact height of a point, enough information is retained to be able to identify types of links. A link is placed in general position with respect to the projection, i.e., no edge of the link is parallel to the projection direction and the projection $\pi : \mathbb{R}^3 \to \mathbb{R}^2$ is regular. $\pi$ is regular if it is injective except for a finite number of points $c_i$ (these are crossings in a link) and only two points in the link are projected onto each $c_i$. Last but not least for each crossing $c_i$ the arc that is on top needs to be distinguished from the one below. The arc passing underneath is usually indicated by a break in the line (so far we have used this convention implicitly).

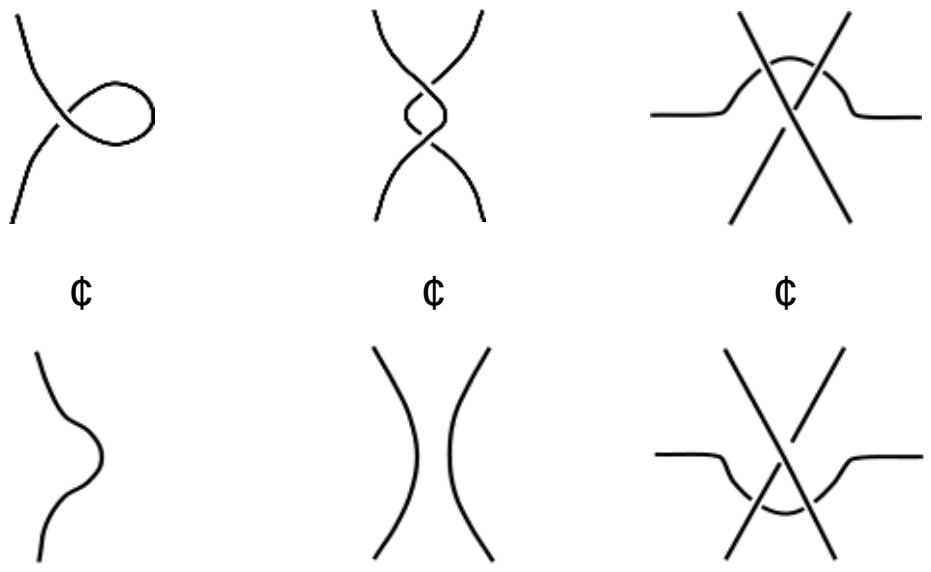

**Figure 5: Reidemeister moves**

Gradually deforming a link that has been projected to two dimensions may lead to a violation of the regularity of the projection $\pi$ at certain step during the transformation. In order to prevent this from happening Reidemeister introduced three types of moves, allowing the arcs of a link to "jump" across critical sections of the transformation [30]; Figure 5 depicts the *Reidemeister moves*. Equivalence of links is then defined as a (finite) sequence of deformations and Reidemeister moves to transform one link into another.

## 3.5 Discussion

The main problem that Grishanov et al. attack with topological methods is the classification of textile structures, i.e., deciding whether two given structures are equivalent [13]. Applying knot theory allows them to re-use results for determining knot equivalence. However, there is a catch: one of the first algorithms by Haken for doing so is extremely complex and was never implemented [15]. Hass et al. review a number of knot algorithms and conclude that they are impractical and that the exact complexity of several general problems in this area is not even clear [16]. Hotz claims to have found an efficient algorithm for the knot equivalence problem [18], but it turns out that the complexity of his algorithm is $O(2^{\frac{n}{3}})$.

Due to the challenge of finding efficient algorithms, the development of *knot invariants* is pursued as well. Invariants are functions mapping the set of knots to some other set (e.g. integers, polynomials, or matrices). Knots from the same equivalence class get mapped to the same element in the codomain of the function. A considerable number of knot and link invariants exist, dividing knots and links into various equivalence classes. One of the simplest invariants for links is the multiplicity $\mu(L)$ of a link $L$, i.e., the number of its components. Some invariants, such as the *unknotting number*, i.e., the minimum number of times a link has to cross itself to be transformed into a trivial link, are simple to formulate, but hard to compute.

In summary, this leaves us with inefficient algorithms of knot equivalence and a comprehensive overview by Grishanov et al. of invariants that can be used to classify doubly-periodic textile structures [13]. However, it is not clear how these techniques can be used for textile pattern retrieval, as no similarity measures between textile structures or invariants are defined. Moreover, the methods often rely on deforming or unknotting an object, which in the context of textile structures would in some cases decompose a fabric into individual strands, unraveling (parts of) the structure. This would make it very difficult to compare the relative location of crossings to each other in two different textiles. Our work, which we start describing in the following section, is inspired by knot theory and especially the concept of tangles, but we follow a different approach taking into account efficiency and performance aspects.

# 4. OUR APPROACH

In the following we first define a hypergraph representation of textile structures, then show how we can extract features, called neighborhoods, from hypergraphs that we can use for measuring the similarity of these graphs.

## 4.1 Textile Graphs

The basic unit we use is a *crossing* of two threads with four links to neighboring crossings. In principle, this is a 2-tangle, but we have exactly one crossing within this structure, breaking down a fabric into its most basic elements. We formalize the representation of textile structures using hypergraphs, which we have chosen due to their better suitability (compared to traditional graphs) to represent objects, their spatial relationships, and contexts [34].

*Definition 1.* A *textile graph* is defined as a hypergraph $H(C, T, \Xi, \Pi, \Omega)$, where $C$ is a set of vertices that belong to crossings, $T$ a set of terminal nodes that end threads, $\Xi$ a set of hyperedges (of degree four) that connect vertices from $C$, $\Pi$ a set of regular edges (of degree two) that indicate which thread is on top in each crossing, and $\Omega$ a set of edges connecting vertices to vertices from other crossings or to terminal nodes.

A textile graph has the following characteristics: $|C|$ is a multiple of four and every $c_i \in C$ belongs to exactly one $x_j \in \Xi$. Every $x_j$ in turn contains only one $p \in \Pi$, that is only two $c_i$ in each crossing are connected via a so-called top edge. In addition to this, we connect every node $c_i$ to either one node from another crossing or one terminal node, i.e., each $c_i$ and each $t_i$ appear in only one $o \in \Omega$. Terminal nodes end threads, so every $t \in T$ has a degree of one. If $T$ is empty, then the structure modeled in the hypergraph represents a knot or link. We can even represent non-textile structures such as chain mail with our approach. Here we focus on textile applications, though.

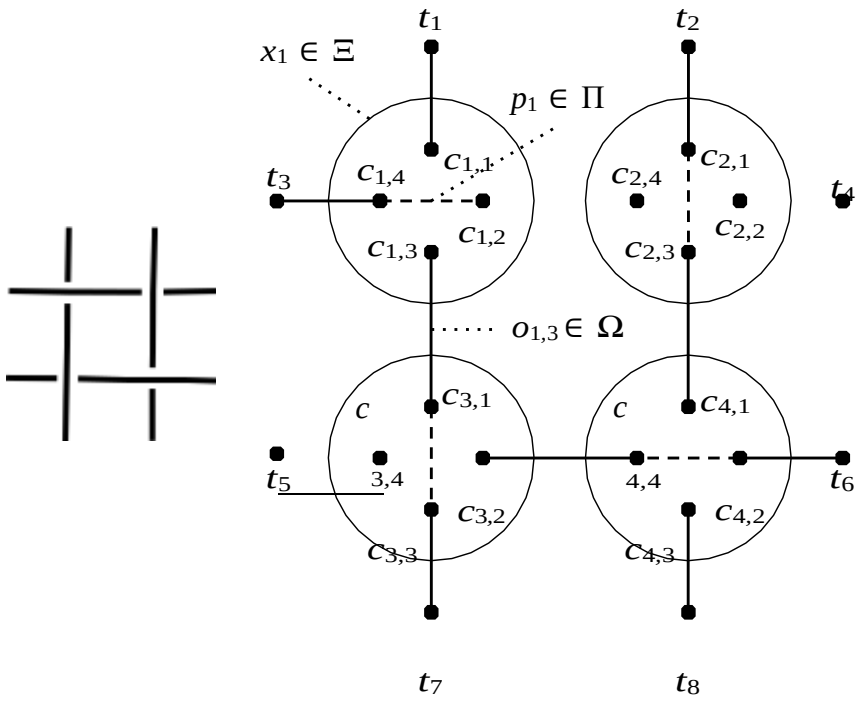

**Figure 6: A textile structure and its hypergraph $H_1$**

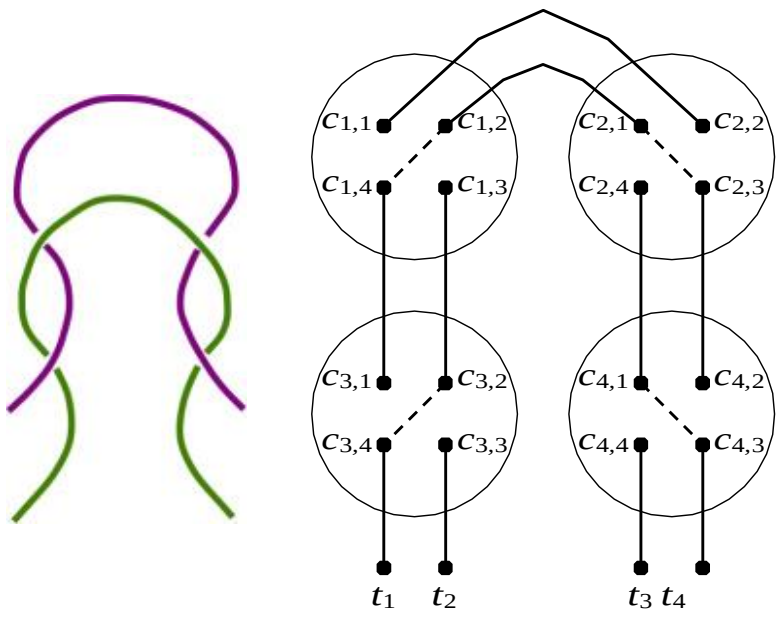

**Figure 7: Another example: hypergraph $H_2$**

Figures 6 and 7 show two examples of simple textile structures and their hypergraph representation. The nodes $c_{i,j}$ belong to $C$, the $t_i$ to $T$. The solid lines are the edges in $\Omega$, the dashed lines those in $\Pi$, while the hyperedges in $\Xi$ are represented by circles.

## 4.2 Comparing Textile Graphs

Now that we have defined textile graphs, we have to formalize the notion of what it means to compare these graphs. One approach is to look for subgraph isomorphisms, which, in the general case, is a hard problem when applied to hypergraphs [2]. While it is interesting to know whether two textile graphs are isomorphic, we have a different goal: we want to measure the similarity of two structures, which means we need a hypergraph similarity measure. While different similarity measures have been proposed for traditional graphs [29], among them graph edit distances (GED) [10], the picture looks different for hypergraphs. There are a few approaches based on mathematical morphology [3] and error-tolerant hypergraph matching, including edit distance [5]. However, common to these approaches is their high computational cost.

We follow an efficient two-phase approach in approximating the similarity of two textile graphs. In a first phase we extract information about the structure of a textile in the form of subgraphs, which we cover in Section 4.3, and in a second phase we measure the similarity based on the extracted information, which is the topic of Section 4.4

## 4.3 Extracting Structural Information

*Star structures*, which are small subgraphs, have been used to describe the internal structure of graphs [35]: every vertex in the graph is interpreted as the root of a tree with depth 2, meaning it includes the vertex itself and all its neighbors. However, this method does not consider any order among the neighbors. While in principle our textile hypergraph is also unordered, we exploit the constraints imposed by the application domain, in particular the fact that we know the relative position of threads in a crossing. (Later on, we also generalize this concept to arbitrary depths, something that Zeng et al. do not do [35].) Before defining the extracted subgraph structure we formalize the positional information of threads, though.

*Definition 2.* Given a node $c_i$ belonging to crossing $x_i \in \Xi$ and a node $c_j$ belonging to crossing $x_j \in \Xi$ ($x_i \neq x_j$) connected by an edge $o_{i,j} = (c_i, c_j) \in \Omega$, we say that $o_{i,j}$ is

- *alternating*, if one of its endpoints is found in $\Pi$ (i.e., it is connected via a top edge) and the other is not. More formally, either $(\exists\, \gamma_i \in x_i : (c_i, \gamma_i) \in \Pi) \wedge (\forall\, \gamma_j \in x_j : (c_j, \gamma_j) \notin \Pi)$ or $(\forall\, \gamma_i \in x_i : (c_i, \gamma_i) \notin \Pi) \wedge (\exists\, \gamma_j \in x_j : (c_j, \gamma_j) \in \Pi)$.
- *non-alternating*, if both of its endpoints are either connected via a top edge or both are not. Formally, $\exists\, \gamma_i \in x_i, \gamma_j \in x_j : (c_i, \gamma_i) \in \Pi \wedge (c_j, \gamma_j) \in \Pi$ or $\forall\, \gamma_i \in x_i, \gamma_j \in x_j : (c_i, \gamma_i) \notin \Pi \wedge (c_j, \gamma_j) \notin \Pi$.
- *terminated*, if one of its endpoints is a terminator: $\exists\, t_j \in T : (c_i, t_j) \in \Omega$

Basically, an alternating thread changes from top to bottom or vice versa from one crossing to the next. For example, the edge from $c_{1,4}$ to $c_{3,1}$ in Figure 7 is alternating, the edge from $c_{1,1}$ to $c_{2,2}$ is not, while $c_{3,4}$ to $t_1$ is terminated.

### 4.3.1 Neighborhoods

The *neighborhood* of a crossing is defined by a 2-tuple containing two sets of labels. The first set specifies whether the outgoing edges from the top edge vertices are alternating ('a'), non-alternating ('n'), or connect to a terminal ('t'). The second set specifies the same for the vertices on the bottom.

*Definition 3.* Given a crossing defined by $x_i \in \Xi$, let $c_{i,1}$ and $c_{i,2}$ stand for the top thread, i.e. $(c_{i,1}, c_{i,2}) \in \Pi$ and $c_{i,3}$ and $c_{i,4}$ for the bottom thread, i.e. $(c_{i,3}, c_{i,4}) \notin \Pi$. $B(x_i) = [\{z_1, z_2\}, \{z_3, z_4\}]$ is the *neighborhood* of crossing $x_i$ where

$$z_j = \begin{cases} \text{'a'} & \text{if } (c_{i,j}, \gamma_{l,j}) \in \Omega \text{ alternating} \\ \text{'n'} & \text{if } (c_{i,j}, \gamma_{l,j}) \in \Omega \text{ non-alternating} \\ \text{'t'} & \text{if } (c_{i,j}, t_l) \in \Omega \text{ terminated} \end{cases}$$

and the $\gamma_{l,j}$ are nodes from the other crossings that the $c_{i,j}$ connect to, or in the case of $t_l$ it is a terminator

For example the neighborhood of crossing 1 in Figure 6 is described by the tuple [{'t', 'a'} { 't', 'a'}]. Hence, one of the main advantages of this approach becomes evident: the representation retains all the relative spatial relationships, but at the same time is orientation invariant. Rotating the textile pattern by 90 or 180 degrees or mirroring the structure has no effect on the textile graph and its crossing neighborhoods.

We describe a textile graph by computing the neighborhood of every crossing in the hypergraph and storing this information as a multiset of tuples.

*Definition 4.* The *fingerprint* $F(H)$ of a textile graph $H(C, T, \Xi, \Pi, \Omega)$ is the multiset of the neighborhoods of its crossings: $F(H) = \{B(x_i) | x_i \in \Xi\}$

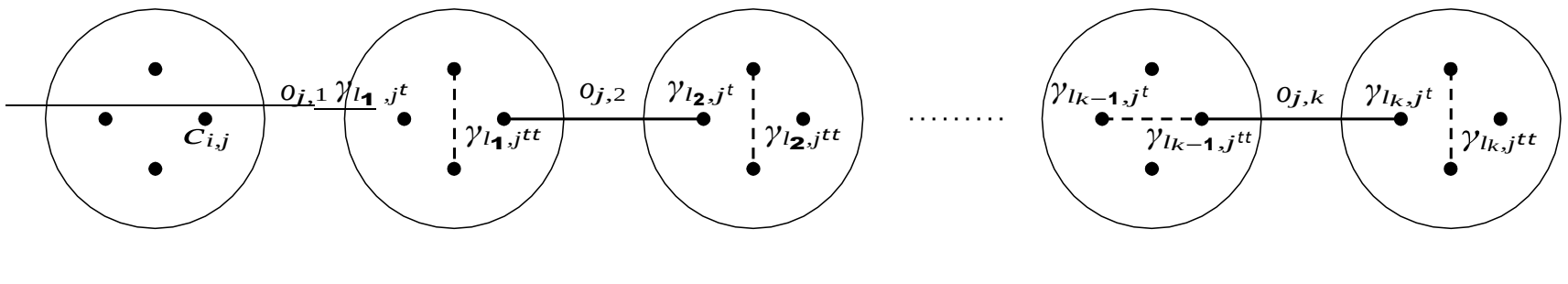

**Figure 8: Illustration of one branch of a** ***k*****-neighborhood**

For example, the fingerprint of the textile graph shown in Figure 6 is $F(H_1) = \{[\{\text{'a', 't'}\}, \{\text{'a', 't'}\}], [\{\text{'a', 't'}\}, \{\text{'a', 't'}\}], [\{\text{'a', 't'}\}, \{\text{'a', 't'}\}], [\{\text{'a', 't'}\}, \{\text{'a', 't'}\}]\}$, which means that all the nodes of the crossings are connected to terminals or are part of alternating edges. This makes sense, as the textile shown in Figure 6 is a plain weave, which is characteristically defined by alternating threads. The fingerprint of the textile in Figure 7, on the other hand, looks different: $F(H_2) = \{[\{\text{'a', 'n'}\}, \{\text{'a', 'n'}\}], [\{\text{'a', 'n'}\}, \{\text{'a', 'n'}\}], [\{\text{'a', 't'}\}, \{\text{'a', 't'}\}], [\{\text{'a', 't'}\}, \{\text{'a', 't'}\}]\}$.

### 4.3.2 k-Neighborhoods

The concept of a neighborhood can be generalized by not just looking at the immediate neighbors of a crossing, but by following a thread to further crossings and checking for each connection whether it is alternating or not. Clearly, we stop tracing a thread as soon as we hit a terminator node. This schema specifies a *k-neighborhood*, in which we follow each of the four emerging threads to the next $k$ neighbors:

*Definition 5.* Given a crossing defined by $x_i \in \Xi$, again let $c_{i,1}$ and $c_{i,2}$ stand for the top thread, i.e. $(c_{i,1}, c_{i,2}) \in \Pi$ and $c_{i,3}$ and $c_{i,4}$ for the bottom thread, i.e. $(c_{i,3}, c_{i,4}) \in \Pi$. Furthermore, let $x_{l_1,j}, x_{l_2,j}, \ldots, x_{l_k,j}$ be the sequence of $k$ crossings we encounter when following the thread leaving $x_i$ via $c_{i,j}$ and let $\gamma_{l_h,j'}, \gamma_{l_h,j''} \in x_{l_h,j}$ be the nodes in each crossing along this thread, i.e., either $\gamma_{l_h,j'}, \gamma_{l_h,j''} \in \Pi$ or $\gamma_{l_h,j'}, \gamma_{l_h,j''} \notin \Pi$. The edges $o_{j,h} \in \Omega$ connect nodes from different crossings, so $o_{j,1}$ connects $c_{i,j}$ and $\gamma_{l_1,j'}$, and for $h \geq 2$ $o_{j,h}$ connects $\gamma_{l_h,j''}$ and $\gamma_{l_h,j'}$. Figure 8 illustrates this situation. Then $B_k(x_i) = [\{y_1, y_2\}, \{y_3, y_4\}]$ is the *k-neighborhood* of crossing $x_i$ with $y_j = [y_{j,1}, y_{j,2}, \ldots, y_{j,k}]$ where each

$$y_{j,m} = \begin{cases} \text{'a'} & \text{if } o_{j,m} \in \Omega \text{ alternating} \\ \text{'n'} & \text{if } o_{j,m} \in \Omega \text{ non-alternating} \\ \text{'t'} & \text{if } o_{j,m} \in \Omega \text{ terminated} \end{cases}$$

If $o_{j,m}$ is a terminated edge for $m < k$, we only have $m$ elements in tuple $y_j$. For the example in Figure 8 the tuple $y_j$ is equal to ['a', 'n', . . . , 'a'].

This makes the neighborhood described in Definition 3 a special case of a $k$-neighborhood with $k = 1$. The 2-neighborhood of crossing $x_1$ in Figure 7, for example, is [ [{'n', 'a'], ['a', 't']}, { ['n', 'a'], ['a', 't']} ]. The fingerprints of hypergraphs using $k$-neighborhoods are computed accordingly, we just have to replace $B(x_i)$ with $B_k(x_i)$ in Definition 4.

## 4.4 Measuring Similarity

Having defined fingerprints of textile patterns, we now have to specify how to actually measure their similarity or distance. Typical distance metrics used for measuring the similarity of objects in non-Euclidean spaces include the Jaccard distance, the Hamming distance, and the cosine measure [19], all of which we cover in the following sections.

### 4.4.1 Jaccard Coefficient

Since the fingerprints of our textile graphs are multisets rather than sets, we need to use the Jaccard coefficient for multisets. For every element in a multiset we store the number of occurrences of the element, so, for example, the multiset $\{a, a, a, b, c, c\}$ becomes $\{a : 3, b : 1, c : 2\}$ or just (3, 1, 2) if we assign fixed positions to each element, position 1 representing the frequency of a, position 2 the frequency of b, and position 3 the frequency of c. Given two multisets $R = (r_1, r_2, \ldots, r_n)$ and $S = (s_1, s_2, \ldots, s_n)$ in a vector representation, $R \cap S$ is computed as $\sum_{i=1}^{n} \min(r_i, s_i)$ and $R \cup S = \sum_{i=1}^{n} \max(r_i, s_i)$. Consequently, we get the following distance measure:

$$D_J(R, S) = 1 - \frac{\sum_{i=1}^{n} \min(r_i, s_i)}{\sum_{i=1}^{n} \max(r_i, s_i)} = \frac{\sum_{i=1}^{n} |r_i - s_i|}{\sum_{i=1}^{n} \max(r_i, s_i)} \quad (1)$$

Computing $D_J(F(H_1), F(H_2))$ gives us the distance between two textile graphs $H_1$ and $H_2$. For example, the fingerprint of $H_1$ (Figure 6) is made up of four times the tuple [{'a', 't'}, {'a', 't'}] and does not contain [{'a', 'n'}, {'a', 'n'}], while the fingerprint of $H_2$ contains both of these tuples two times each, i.e., we can represent $H_1$ by $(0, 4)^T$ and $H_2$ by $(2, 2)^T$. Applying Formula (1), we obtain $\frac{2+2}{2+4} = \frac{2}{3}$.

### 4.4.2 Hamming Distance

Fixing the positions of the elements within the multisets allows us to interpret them as vectors. Using a vector representation, we can apply the Hamming distance, which computes the number of components that differ in two vectors. Let $f_H(r_i, s_i)$ be a function that compares two components, such that

$$f_H(r_i, s_i) = \delta_{r_i, s_i} = \begin{cases} 0 & \text{if } r_i = s_i \\ 1 & \text{if } r_i \neq s_i \end{cases}$$

Then a formal definition of the Hamming distance yields

$$D_H(R, S) = \sum_{i=1}^{n} f_H(r_i, s_i) \quad (2)$$

Applying Formula (2) to the frequency vectors of the fingerprints of $H_1 = (0, 4)^T$ and $H_2 = (2, 2)^T$ gives us 1 + 1 = 2.

There are a few issues with the Hamming distance. First, it is not normalized, the distance between two vectors ranges from 0 to $n$, which even varies depending on the size of the vectors. Second, if the elements of the domain used for the vector components are comparable, e.g. in the case of integers, intuitively the vector $(1, 0, 3)^T$ is closer to $(1, 0, 2)^T$ than $(1, 0, 7)^T$. However, applying the original definition would give us a distance of 1 in both cases. We can redefine the function $f_H$ to consider this fact: $f_H(r_i, s_i) = |r_i - s_i|$. Consequently, the distance measure becomes

$$D_{\bar{H}}(R,S) = \sum_{i=1}^{n} f_{\bar{H}}(r_i, s_i) = \sum_{i=1}^{n} |r_i - s_i| \quad (3)$$

Applying Formula (3) to the frequency vectors of the fingerprints of $H_1 = (0, 4)^{\mathsf{T}}$ and $H_2 = (2, 2)^{\mathsf{T}}$ gives us 2 + 2 = 4.

### 4.4.3 Cosine Measure

Formally, the cosine distance measure is defined via an inner vector product:

$$D_c(R, S) = 1 - \frac{\sum_{i=1}^{n} r_i \cdot s_i}{\sqrt{\sum_{i=1}^{n} r_i^2} \cdot \sqrt{\sum_{i=1}^{n} s_i^2}} \quad (4)$$

$D_c(F(H_1), F(H_2))$ gives us the cosine measure distance between two textile graphs $H_1$ and $H_2$. For example, applying Formula (4) to the frequency vectors of the fingerprints of $H_1 = (0,4)^{\mathsf{T}}$ and $H_2 = (2,2)^{\mathsf{T}}$ yields $1 - \frac{0+8}{\sqrt{4}\sqrt{8}} = 1 - \sqrt{2}/2$.

Often term frequency (TF) and inverse document frequency (IDF) are applied to the vectors when using the cosine measure. This is done to consider the fact that an increase in term frequency within an individual document has less and less impact and that terms appearing less often in a document collection tend to be more important. In our case we also compare textiles using logarithmic TF-IDF factors: $\mathrm{TF}_{p,t} = 1 + \log(f_{p,t})$ and $\mathrm{IDF}_p = \log\frac{N}{f_p}$ where $f_{p,t}$ is the frequency of fingerprint $p$ in textile $t$, $N$ is the overall number of textiles in the collection, and $f_p$ is the number of textiles in the collection in which fingerprint $p$ occurs.

## 5. IMPLEMENTATION DETAILS

Algorithm 1 shows a description of the fingerprint computation in pseudo-code. We iterate through all crossings and select all four nodes of a crossing in turn to follow a thread to the next $k$ crossings, noting any change of position on the way. Let the first two nodes, $c_{i,1}$ and $c_{i,2}$, denote those of the top-level thread, i.e., $(c_{i,1}, c_{i,2}) \in \Pi$, while the other two nodes, $c_{i,3}$ and $c_{i,4}$, belong to the bottom-level thread, i.e., $(c_{i,3}, c_{i,4}) \notin \Pi$. We use the function NEIGHBOR to find the connecting node in a neighboring crossing, the function FINDLABEL to determine the label of an edge, and the function OPPOSITE to find a node's counterpart within a crossing. If we reach the end of a thread before encountering $k$ crossings, we fill up the labels with NULL values.

We have implemented our algorithm efficiently by storing the crossing nodes of a hypergraph in an array. Every node has the following structure:

```
struct node {
  int  nextNode;
  bool onTop;
  int  oppositeNode;
}
```

The four nodes of a crossing are stored in neighboring cells of the array, i.e., the nodes at positions $4i$ to $4i + 3$ belong to crossing $i$ (for $0 \leq i \leq n - 1$, assuming we have $n$ crossings). The component `nextNode` connects a node to the node of a neighboring crossing by storing the index of this node. If a node connects to a terminal, `nextNode` is equal to -1, which means that we do not have to store the terminal nodes explicitly. In order to determine whether a node belongs to a top thread, we have added the Boolean variable `onTop`. The component `oppositeNode` is not strictly necessary, as we could determine the opposite node by looking at all other nodes of a crossing and then selecting the one with the same value for `onTop`; it was added for efficiency reasons.

**Algorithm 1:** FINGERPRINT($H$,$k$)

```
Input : hypergraph H(C, T, Ξ, Π, Ω) with
          C: set of vertices belonging to crossings
          T : set of terminals
          Ξ: set of hyperedges connecting vertices
          Π: set of edges indicating top thread
          Ω: set of edges connecting crossings/terminals
          k: size of the neighborhood
Output: fingerprint for hypergraph H
FP := ∅;
for every x_i ∈ Ξ do
    for j := 1 to 4 do
        let c_{i,j} be the j-th node of crossing x_i;
        (* j = 1, 2 for top-level thread *)
        l := 0, current = c_{i,j};
        repeat
            l++;
            γ_{l,jt} := NEIGHBOR(current);
            label_{j,l} := FINDLABEL(current, γ_{l,jt});
            if label_{j,l} ≠ 't' then
                γ_{l,jtt} := OPPOSITE(γ_{l,jt});
                current := γ_{l,jtt};
            end
        until k = l or label_{j,l} = 't';
        for s := l + 1 to k do
            label_{j,s} := NULL;
        end
    end
    fp_i := [{[label_{1,1}, label_{1,2}, . . . , label_{1,k}],
             [label_{2,1}, label_{2,2}, . . . , label_{2,k}]},
            {[label_{3,1}, label_{3,2}, . . . , label_{3,k}],
             [label_{4,1}, label_{4,2}, . . . , label_{4,k}]}];
    FP := FP ∪ fp_i;
end
return FP;
```

In summary this means that the $k$-neighborhoods of all $n$ crossings of a textile can be computed in $O(nk)$.

## 6. EXPERIMENTAL EVALUATION

We evaluated the variants of our similarity measure experimentally, clustering a data set containing over a thousand different textiles and comparing the outcome to the correct classification. Additionally, we also run queries over our data set, measuring the retrieval performance. We also look at the efficiency, presenting numbers on the run time of the algorithm.

### 6.1 Experimental Setup

The algorithm was implemented using Java JDK 1.7.0_60 running under Windows 7. All experiments were run on a computer with an Intel Core i5 CPU (2.60 GHz) and 2 GB memory.

In order to test the effectiveness of our textile similarity measures, we partition a collection of $n$ textile structures represented as hypergraphs $S = \{H_1, H_2, \ldots, H_n\}$ into $m$ clusters $L = \{\lambda_1, \lambda_2, \ldots, \lambda_m\}$ and compare the outcome to the $m$ categories of the correct classification $A = \{\alpha_1, \alpha_2, \ldots, \alpha_m\}$.

For clustering, we use hierarchical agglomerative clustering, in which each textile $H_i$ starts out in its own cluster. In every sub-sequent step, the two nearest subclusters are merged until $m$ clusters remain. The Unweighted Pair Group Method with Arithmetic Mean (UPGMA) is used to calculate the distance between two subclusters $u_i$ and $u_j$:

$$D_u(u_i, u_j) = \frac{1}{|u_i||u_j|} \sum_{H_r \in u_i} \sum_{H_s \in u_j} D(F(H_r), F(H_s)) \quad (5)$$

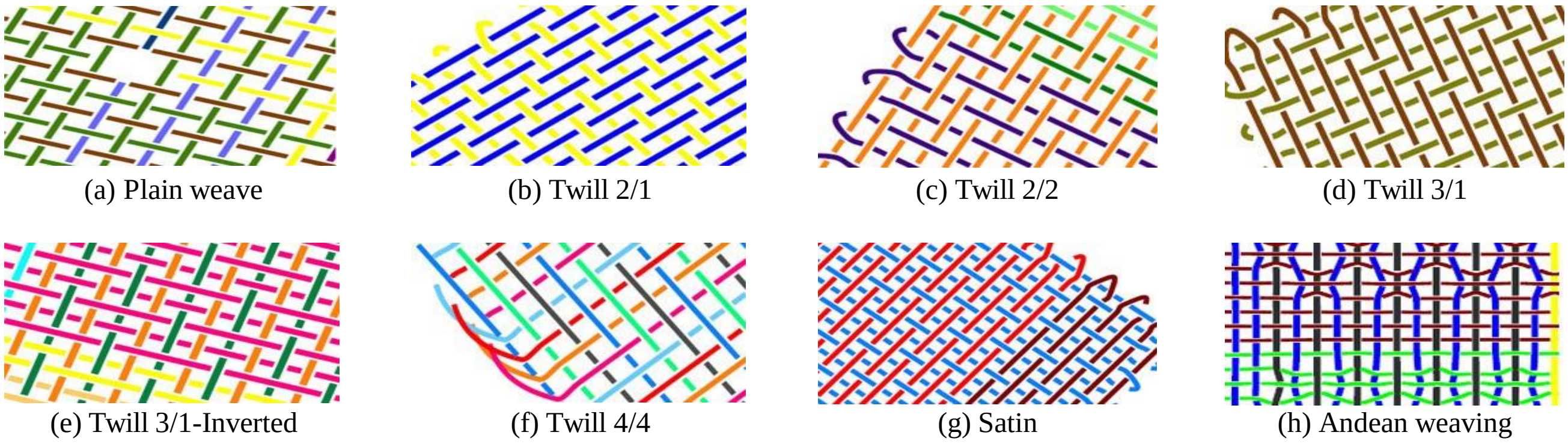

(a) Plain weave (b) Twill 2/1 (c) Twill 2/2 (d) Twill 3/1

(e) Twill 3/1-Inverted (f) Twill 4/4 (g) Satin (h) Andean weaving

**Figure 9: Examples of weaving patterns**

For the purpose of measuring the quality of the clustering $L$ compared to the correct classification $A$, we apply the Rand index [28]. Every pair of textiles $H_i, H_j \in S, i > j$ is categorized as a true positive (TP), true negative (TN), false positive (FP), or false negative (FN), depending on the following conditions:

TP: $H_i$ and $H_j$ are in the same cluster in $L$ and in the same class in $A$.

TN: $H_i$ and $H_j$ are in different clusters in $L$ and in different classes in $A$.

FP: $H_i$ and $H_j$ are in the same cluster in $L$ and in different classes in $A$.

FN: $H_i$ and $H_j$ are in different clusters in $L$ and in the same class in $A$.

The Rand index (RI) then measures the ratio of textiles placed correctly within the clustering:

$$RI = \frac{TP + TN}{TP + FP + TN + FN}$$

The Rand index assigns equal weight to false positives and false negatives and also includes true negatives. Usually, it is not very difficult to identify a large number of true negatives correctly, therefore we also use the standard quality measures of precision $P = \frac{TP}{TP+FP}$, recall $R = \frac{TP}{TP+FN}$, and F-measure $F = \frac{2PR}{P+R}$.

We evaluate the retrieval performance by using each of the textile objects as a query $q_j \in Q$ and then ranking all the other textiles according to their similarity to the query. All the textiles $\{h_1, h_2, \ldots, h_{m_j}\}$ that are in the same category $\alpha_j$ as $q_j$ are considered to be relevant, while those from other categories are not relevant. We measure the quality of the resulting ranked lists using mean average precision (MAP), average Precision-Recall (PR), and average F-measure-Recall (FR) curves [23]. MAP provides a single value measuring the quality across all recall levels:

$$MAP(Q) = \frac{1}{|Q|} \sum_{j=1}^{|Q|} \frac{1}{m_j} \sum_{k=1}^{m_j} Precision(R_{jk})$$

where $R_{jk}$ is the list of from the first textile object down to $h_k$ and $Precision(R_{jk})$ is the precision of the set $R_{jk}$.

A PR curve shows the change of precision with increasing recall; we calculate the standard 11-point interpolated average precision. The interpolated precision of query $q_j$ at the standard recall level $r_l$, $0 \leq l \leq 10$, is defined as the highest precision found for any recall level $r \geq r_l$: $P_i(r_l) = \max_{r \geq r_l} P_i(r)$, where $P(r)$ is the precision at recall level $r$. Therefore, the average precision of $Q$ at the standard recall level $r_l$ is defined as:

$$\bar{P}(r_l) = \frac{\sum_{i=1}^{|Q|} P_i(r_l)}{|Q|}$$

Similarly, the average F-measure of $Q$ at the standard recall level $r_l$ is equal to:

$$\bar{F}(r_l) = \frac{\sum_{i=1}^{|Q|} F_i(r_l)}{|Q|}$$

where $F_i(r_l) = \frac{2P(r_l)r_l}{P(r_l)+r_l}$ is the interpolated F-measure of query $q_j$ at $r_l$.

## 6.2 Data Set

We employ a textile editor called SAWU [14, 24] to model the textiles in the data set and export their representations into plain text files, which we use to create the corresponding hypergraphs. With the help of a domain expert the 1200 fabrics were divided into twelve categories, each containing 100 items of a particular type of textile. On average, each textile consists of 25,352 vertices, 6,311 crossings (hyperedges) and 12,679 edges connecting crossings with each other or with terminals.

In the following, we give an overview of the different kinds of textiles found in each group. One of the simplest weaving patterns is plain weave, in which a weft thread alternates between going over and under a warp thread.[2] In each row, this pattern is shifted by one position (see Figure 9(a)). The next five groups of patterns consist of twills, in which more than one warp thread is crossed over or under. Figures 9(b) to (f) show example patterns, ranging from 2/1 twill to 4/4 twill. In the satin (also known as sateen) weave structure (see Figure 9(g)), four or even more weft threads float over a warp thread or vice versa. The most complex patterns in our collection are taken from a collection of weavings originating in the Andes. Since they were created manually, they can exhibit a great variety of different styles in a single textile. The pattern depicted in Figure 9(h) indicates this, as the warp and weft threads cross a different number of threads in different parts of the textile. For more examples of Andean textiles, please see `http://www.weavingcommunities.org/`.

[2]Warp threads are longitudinal threads held in place by a frame, while the weft thread is led through the warp threads.

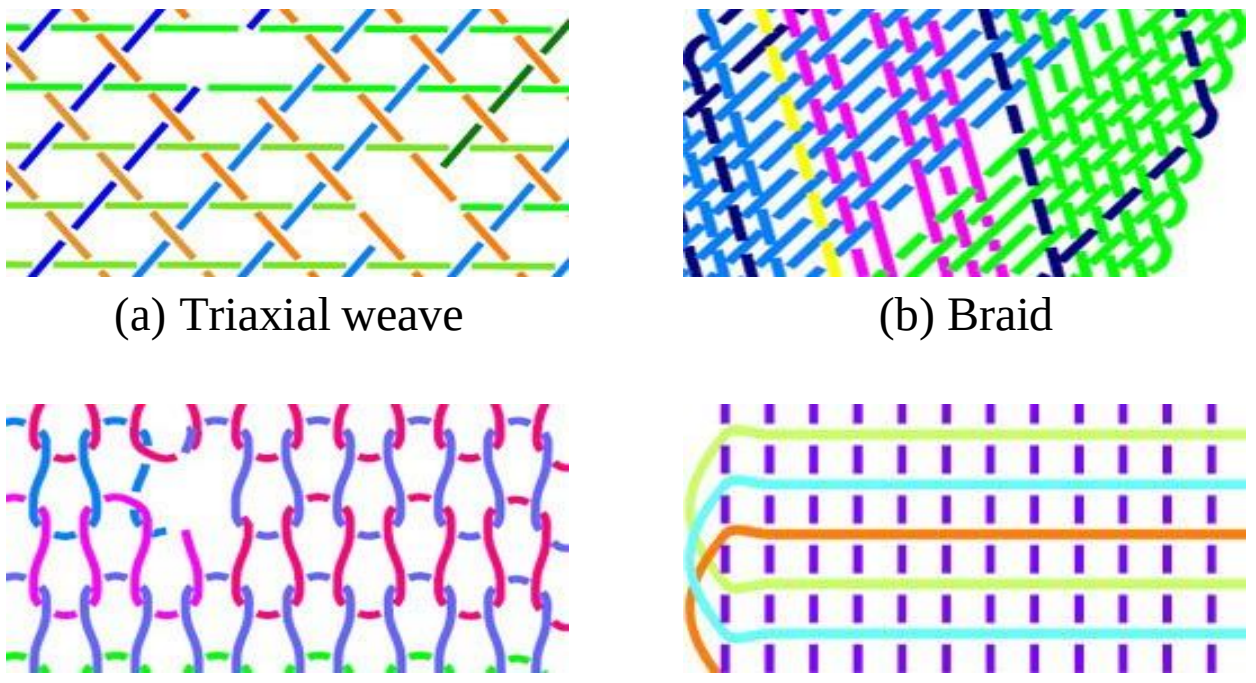

(a) Triaxial weave (b) Braid

(c) Knitting (d) Warp above weave

**Figure 10: More examples of textile patterns**

For the remaining groups of textiles, shown in Figure 10 we have chosen patterns that are not actually woven. Triaxial weave, although called a weave, is a hybrid structure between weaving and braiding. The resulting structure, an example of which can be seen in Figure 10(a), does not follow a rectilinear pattern. Braids are created by intertwining three or more threads as shown in Figure 10(b). In knitting (see Figure 10(c)), loops are formed by connecting a row of new loops to a row of already existing loops. When done manually, this usually involves needles holding the thread. In the warp above weave pattern all the threads of one type are always located above the other (see Figure 10(d)). We included this group to see how our similarity measure would cope with non-textile patterns.

We have also introduced imperfections into some of the textiles in each group to test the similarity measure's capability to deal with errors in a pattern. Additionally, we also rotated and mirrored some of the textile samples to check that our similarity measure can cope with differently oriented versions of the same weaving pattern.

## 6.3 Results

Basically, we have two parameters with which we can calibrate our model: the size $k$ of the neighborhoods and the distance metric used for comparing two fingerprints (JaccardDist, HammingBool, HammingFreq, CosineFreq, and CosineTfIdf; see Section 4.4 for details). In the following we investigate the impact of both parameters on the run time and on the cluster performance of our algorithms.

### 6.3.1 *Run Time*

Figure 11 illustrates how the run time varies with increasing $k$ for the different distance metrics. Every data point in Figure 11 averages the execution time of nine runs each generating a complete distance matrix including the results for the pairwise comparisons of all textiles. In general, the run time of each variant of our algorithm increases linearly with $k$.

Unsurprisingly, HammingBool, being the simplest formula, is fastest. HammingFreq is slower than HammingBool and slightly faster than JaccardDist, as JaccardDist needs to normalize its result. Slightly surprising is the speed with which the cosine measure variants run, as this involves the most complicated computations compared to the other metrics. This is due to our implementation of the multisets. We refrained from using an explicit vector representation because of the sparsity of the vectors. For instance, although there are 97,871 different (potential) neighborhoods for $k = 9$, on average only 668 appear in a given textile structure. As a consequence we can skip a large part of the inner product calculation.

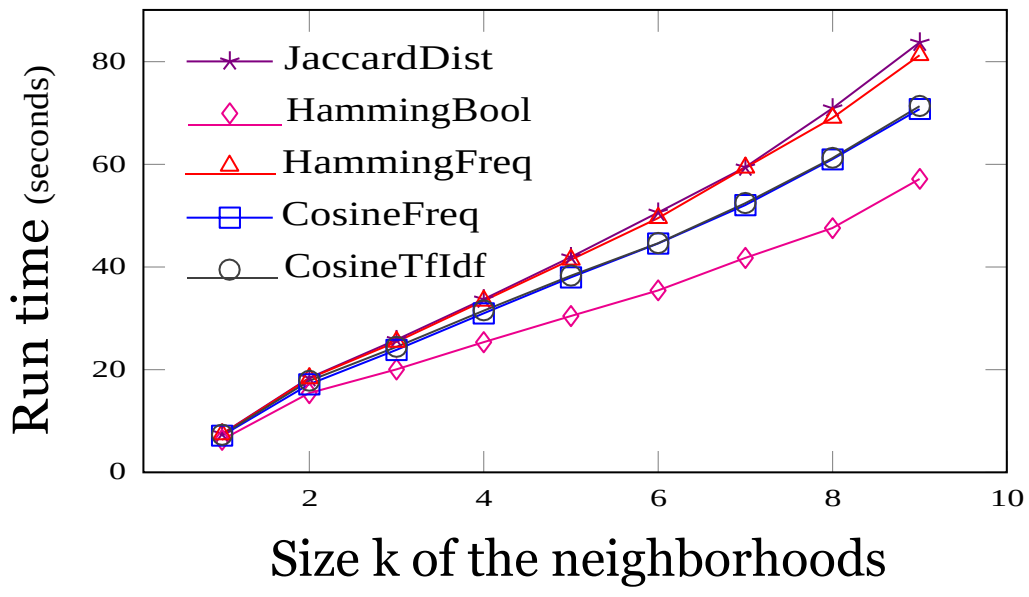

**Figure 11: Run Time**

### 6.3.2 *Cluster Performance*

Figure 12 shows the Rand index, while Figure 13 presents the results for precision, recall, and F-measure for the cluster performance. For HammingBool every $k$-neighborhood has the same influence, regardless of its frequency. This means, that an erroneous neighborhood (due, for example, to imperfections in the textile) will have the same impact as hundreds or thousands of correct identical ones. Additionally, the larger the $k$, the more neighborhoods will be affected, which leads to the peculiar results seen in Figures 12 and 13: the performance of HammingBool becomes worse with increasing $k$. To a lesser extent, this is also true for HammingFreq, although storing the frequency of each neighborhood almost compensates for the negative impact of a growing $k$.

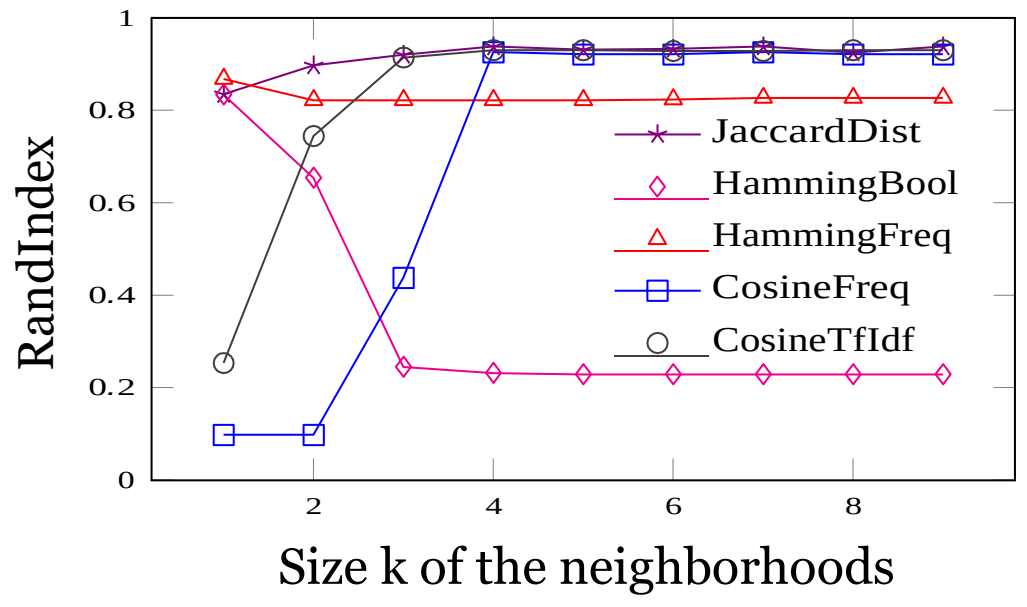

**Figure 12: Rand Index**

In general, the cosine measure performs in a more expected way. When increasing $k$, the precision goes up at the price of a slightly decreasing recall. However, the cosine measure variants have a very low precision for small values of $k$ ($k < 4$), making it much worse than JaccardDist for these cases. We found that the cosine measure tends to underestimate the distance between different objects for small values of $k$, meaning that the wrong patterns are clustered together, leading to more false positives and therefore lower precision. JaccardDist shows the most stable behavior of all the investigated distance metrics, while exhibiting a good cluster performance, so we recommend using it over the others. Also, looking at the results in a general way, we can see that using neigh-borhoods with $k > 4$ does not bring any substantial improvements, on the contrary, sometimes we observed a slightly worse performance.

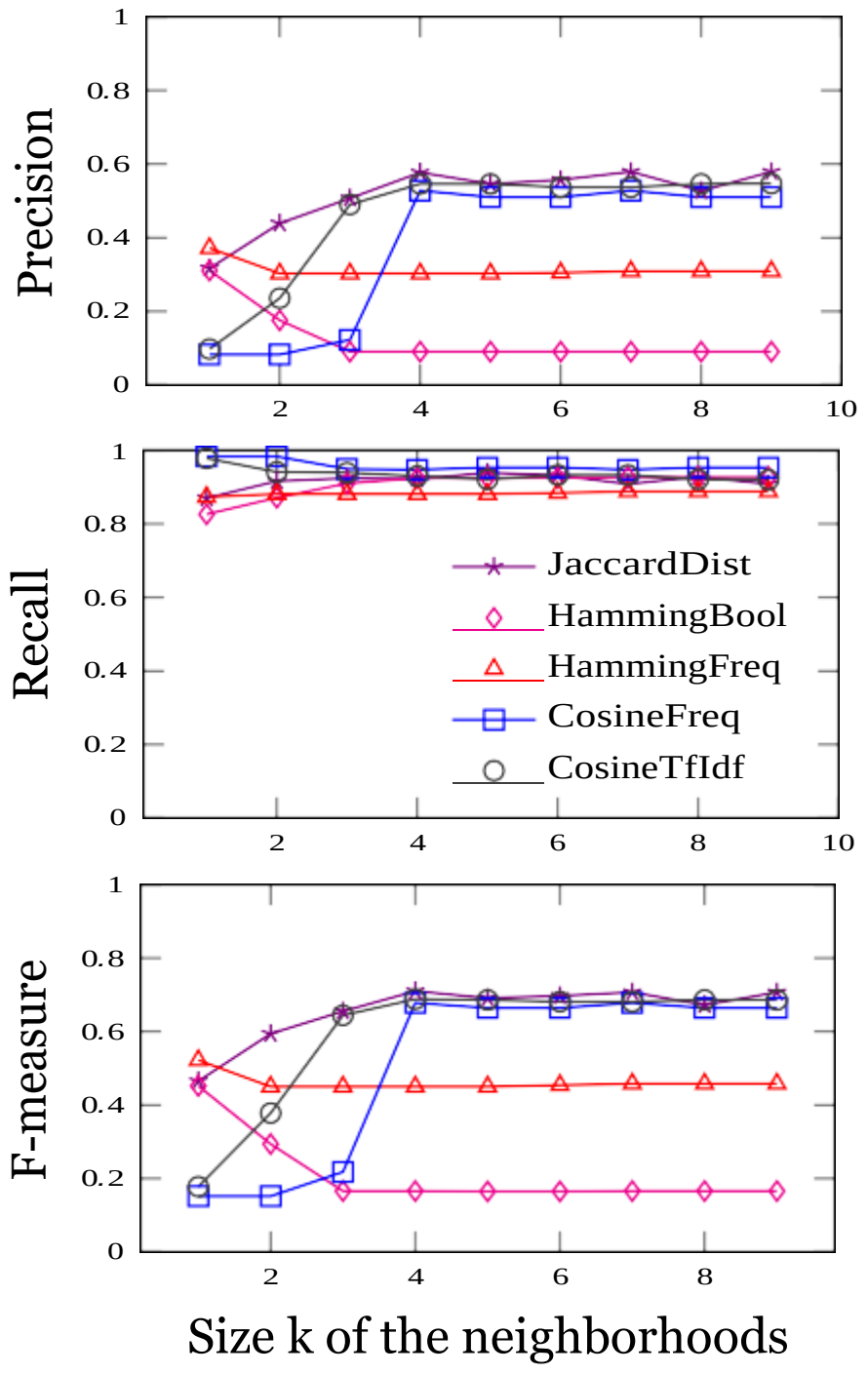

**Figure 13: The other measures**

### 6.3.3 *Retrieval Performance*

Figure 14 depicts the mean average precision (MAP) of the different techniques and indicate the overall utility of our textile pattern similarity measure. As for the cluster performance, there is no significant gain in using neighborhoods with a size greater than four, on the contrary, for some techniques the performance even gets worse. The Jaccard distance variant is clearly on top. For the other variants, the situation is not that clear: HammingFreq and CosineTfIdf are roughly comparable (except for very small values of $k$), while HammingBool and CosineFreq trade places at $k$ equal to four.

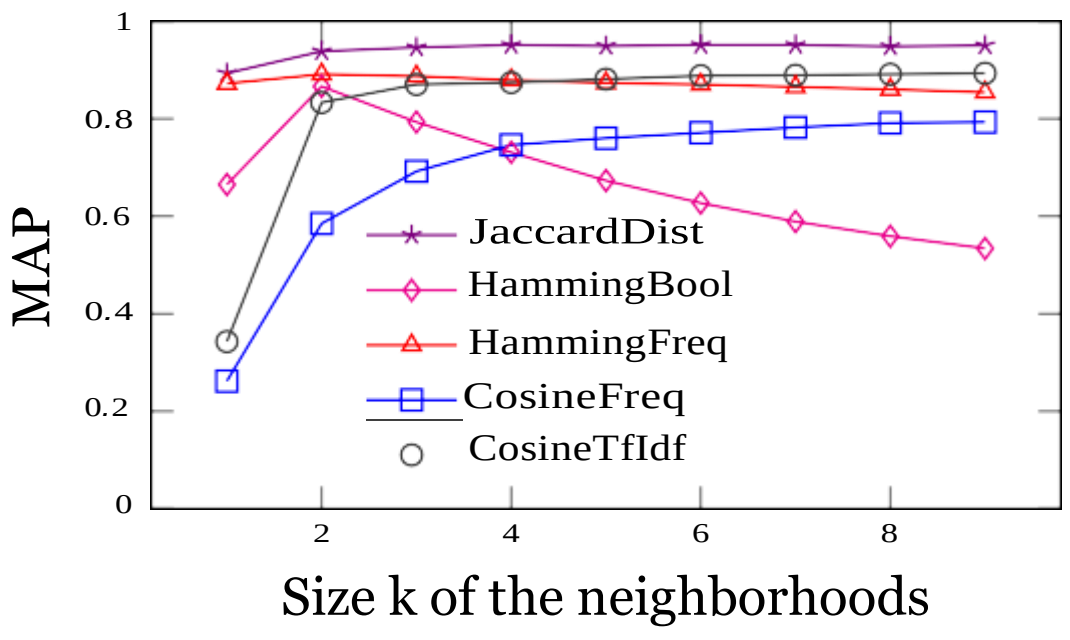

**Figure 14: Mean Average Precision**

Figure 15 shows the PR and FR curves for neighborhoods of size four. Again, the Jaccard distance shows excellent results, in the PR curve the precision stays above 90% for recall values up to 90% and then drops to around 75%. HammingBool steadily loses ground, while HammingFreq is able to keep up with CosineTfIdf. CosineFreq loses about 20% precision within the first 10% of recall and then is able to keep this level, but is not competitive compared to some of the other techniques. Although the gaps in performance between the different similarity measures are smaller, the FR curve shows a similar picture.

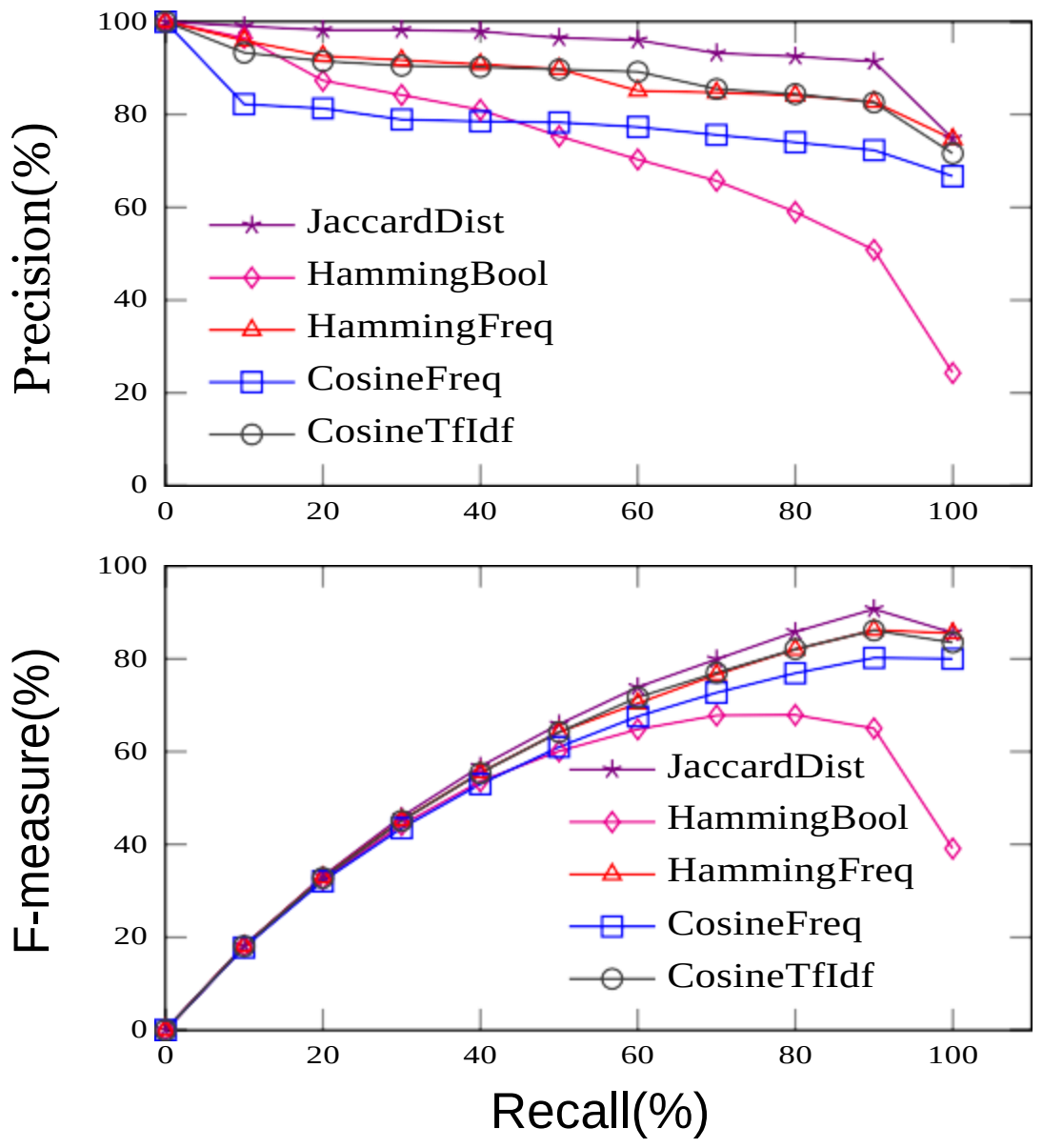

**Figure 15: Average PR and FR curves (4-neighborhoods)**

## 7. CONCLUSION AND FUTURE WORK

More and more information is made available in a digitized form, this also includes areas such as the humanities and cultural heritage. A clear advantage of this approach is that digitized material is much easier to access. However, this also brings new challenges with it: users want to be able to search collections in an adequate and fast way. While text search is fairly well understood, other areas still have a lot of catching up to do. For instance, digital archives for textiles offer keyword search and arrange their content taxonomically, but search functionality on the level of thread structure is still missing.

We developed a technique based on hypergraphs to represent textiles using a crossing of two threads as the basic building block. Decomposing such a graph into substructures called $k$-neighborhoods allows us to determine the similarity of the patterns created by the interwoven threads. In turn, this makes it possible to search a collection of textile patterns given a query pattern. We have experimentally tested several distance metrics for computing the similarity between multisets of $k$-neighborhoods and found that all of them can be implemented efficiently. Nevertheless, when it comes to actually distinguishing between differently made textiles, there are huge differences. The best variants are Jaccard distance and cosine measure, which are able to cluster textiles taken from a data set of 1200 samples much better than the Hamming distance. We also ran queries over the data set, measuring the retrieval performance. Here, the cosine measure loses some ground compared to the Hamming distance, but Jaccard still comes out on top. Consequently, we recommend using Jaccard, as it shows the best overall performance.

For future work we would like to investigate further distance measures and variations of neighborhoods to see if the technique using *k*-neighborhoods combined with the Jaccard distance can be further improved on. Defining a notion of edit distances on hypergraphs for textile structures also looks like a promising direction to take. Finally, evaluating our similarity measure on other data sets to show that it is universally applicable is also an important point to consider. However, at the moment the modeling of the textiles used for the hypergraph representation has to be done manually, in order to automate this process, image-processing techniques for extracting a thread structure and mapping it to graphs would be an interesting topic to look into.